\documentclass[12pt]{article}
\usepackage[english]{babel}
\usepackage{amsmath,amssymb}
\usepackage[dvips]{graphicx}

\def\rpv{$R_p \hspace{-1em}/\;\:\hspace{0.2em}$}
\def\rpvx{$R_p \hspace{-1em}/\;\:$ }
\def\lsim{\raise0.3ex\hbox{$\;<$\kern-0.75em\raise-1.1ex\hbox{$\sim\;$}}}

\newcommand{\AddrAHEP}{%
  AHEP Group, Institut de F\'{\i}sica Corpuscular --
  C.S.I.C. Val{\`e}ncia \\
  Edificio Institutos de Paterna, Apt 22085, E--46071 Valencia, Spain}

\begin{document}

\begin{flushright} \small
hep-ph/0606061 \\
IFIC/06-20 
\end{flushright}

\begin{center}
{\Large R-parity violation: Hide \& Seek} \\[1cm]

M.~Hirsch\footnote{mahirsch@ific.uv.es} and
W.~Porod\footnote{porod@ific.uv.es} \\[2mm] \AddrAHEP \\[5mm]
6$^{th}$ June 2006

\end{center}


\begin{abstract}
We point out that, if R-parity is broken spontaneously, the neutralino 
can decay to the final state majoron plus neutrino, which from the 
experimental point of view is indistinguishable from the standard
missing momentum 
signal of supersymmetry. We identify the regions of parameter space 
where this decay mode is dominant and show that they are independent 
of R-parity conserving SUSY parameters. Thus, (a) only very weak limits 
on R-parity violating couplings can be derived from the observation of 
missing momentum events and (b) at future collider experiments huge 
statistics might be necessary to establish that R-parity indeed 
is broken. Parameter combinations which give calculated relic neutralino 
density larger than the measured dark matter density in case of 
conserved R-parity are valid points in this scenario and their 
phenomenology at the LHC deserves to be studied. 

\end{abstract}



Is R-parity broken? Undoubtedly, a definite, affirmative answer to this 
question would have profound implications for both, particle physics and 
cosmology. Low-energy precision data can be used to derive limits on 
R-parity violating (\rpvx) couplings \cite{Dreiner:1997uz}, but 
experimental proof that R-parity indeed is broken can only come from 
the observation of a decaying lightest supersymmetric particle (LSP) 
at a future collider experiment.

In this paper we present and discuss a scenario with spontaneously
broken R-parity where the model can be easily confused with the usual
Minimal Supersymmetric Standard Model (MSSM) with conserved R-parity
despite the fact that the LSP decays within a typical detector of
present and future high energy experiments. We concentrate here on
neutral LSPs as this is the case which has been studied mostly by
experimental groups. However, in principle the LSP could be charged
and quasi-stable on collider time scales and, thus, be observable.

Suppose supersymmetric particles exist with masses, such that they can 
be produced at the LHC. To obtain evidence that the LSP decays, at 
least the following two conditions need to be met: (i) The decay length 
should not be (much) larger than the size of a typical detector. And, 
(ii) a minimum percentage of LSP decays, the exact number depending on 
the SUSY spectrum and the luminosity of the collider, of course, should 
lead to visible final states. It is straightforward to estimate that 
neutralino ($\tilde\chi^0$) LSP decay lengths, for typical SUSY masses 
in the vicinity of ${\cal O}$(100) GeV, drop below $L \sim 1$ m for \rpv 
couplings larger than - very roughly - ${\cal O}(10^{-6})$. Condition 
(ii), however, is more subtle. Consider the lepton number violating part 
\footnote{Spontaneous \rpv conserves baryon number. One can also use 
baryon parity, or any similar symmetry, to get rid of the dangerous 
baryon number violating terms, see \cite{Ibanez:1991pr}.}  
of the {\em explicit} \rpv superpotential \cite{HallSuzuki}
\begin{equation} \label{spot}
{\cal W} = \epsilon_i \widehat L_i \widehat H_u + 
           \lambda_{ijk}\widehat L_i\widehat L_j \widehat E^c_k +
           \lambda'_{ijk}\widehat L_i\widehat Q_j \widehat D^c_k 
\end{equation}
together with the corresponding bilinear soft terms,
\begin{equation} \label{vsoft}
V_{\rm soft} = \epsilon_i B_i {\tilde L}_i H_u. 
\end{equation}
For trilinear \rpvx, final states of $\tilde\chi^0$ decays always involve 
either two quarks and/or charged leptons. In bilinear \rpvx, scalar 
neutrino vacuum expectations values (vevs) induce couplings such as 
$C_{\tilde\chi^0Z^0\nu}$ and $C_{{\tilde \nu}\nu\nu}$, which lead to the decay 
$\tilde\chi^0 \rightarrow 3 \nu$. Numerically \cite{Porod:2000hv}, however, 
the branching ratio of this mode is always Br($\tilde\chi^0 \rightarrow 3\nu$) 
$< 0.1$, since (a) Br$(Z^0 \rightarrow {\rm inv}) \simeq 0.2$ and 
(b) the coupling $C_{\tilde\chi^0Z^0\nu}$ necessarily is 
$C_{\tilde\chi^0Z^0\nu} \simeq C_{\tilde\chi^0W^{\pm}l^{\mp}}$. Thus, in 
explicit \rpv condition (ii) is automatically fulfilled. 

{\em Spontaneous} violation of R-parity (s-\rpv) 
\cite{Aulakh:1982yn,Masiero:1990uj}, on the other hand, 
implies the existence of a Goldstone boson, usually called the majoron 
(J). In s-\rpv the neutralino can then decay according to $\tilde\chi^0 
\rightarrow J \nu$, i.e.~completely invisible. Below we will show 
that this decay mode can in fact easily be the dominant one, with 
branching ratios very close to 100 \%. From the point of view of 
accelerator phenomenology, models with spontaneous violation 
of R-parity thus can resemble the MSSM 
with conserved R-parity and huge statistics might be necessary 
before it can be established that R-parity indeed is broken. 

Before turning to the details of the calculation, two more comments are 
 in order. Firstly, experiments on solar and atmospheric neutrinos have 
shown that neutrinos have non-zero masses and mix \cite{Fukuda:1998mi}. 
Many models have been proposed, which potentially can explain these data. 
However, the presence of lepton number violation in \rpv inevitably 
implies Majorana neutrino masses. Thus, significant {\em upper} bounds on most 
\rpv parameters follow from oscillation experiments \cite{Rakshit:1998kd}
as \rpv naturally implies a hierarchical neutrino mass spectrum.
On the other hand, even the simplest bilinear \rpv model can explain 
measured neutrino masses \cite{hirsch:2000ef}, as can s-\rpv 
\cite{Hirsch:2004rw}, if some specific combinations of \rpv parameters 
obey certain {\em lower} bounds, see below. In our numerical calculations 
we will always use \rpv parameters fitted to neutrino physics. A short 
comment on how our results are affected by other choices of \rpv parameters 
is given near the end of this paper.

Secondly, recent WMAP data \cite{Spergel:2006hy} have confirmed the 
existence of non-baryonic dark matter and measured its contribution 
to the energy budget of the universe with unprecedented accuracy. 
Because the neutralino as a dark matter candidate is lost, most 
advocates of SUSY declare \rpv undesirable. However, this does not
mean that in models with \rpvx, the dark matter problem could not 
be solved. Possible DM candidates in \rpv are (i) light gravitinos 
\cite{Takayama:2000uz,Hirsch:2005ag}, (ii) the axion 
\cite{Kim:1986ax,Raffelt:1996wa} or (iii) its superpartner, the axino 
\cite{axino}, to mention a few.

Turning to the definition of our model, we note that early versions of 
spontaneous \rpvx used the left sneutrinos to break lepton number 
\cite{Aulakh:1982yn}. These models are now ruled out by LEP and 
astrophysical data \cite{Raffelt:1996wa,lep}. The model we consider 
\cite{Masiero:1990uj} therefore extends the MSSM by three additional 
singlet superfields, namely, $\widehat\nu^c$, $\widehat S$ and $\widehat\Phi$, 
with lepton number assignments of $L=-1,1,0$ respectively. The 
superpotential can be written as \cite{Masiero:1990uj} 
\begin{eqnarray} %
{\cal W} &=& h_U^{ij}\widehat Q_i \widehat U_j\widehat H_u 
          +  h_D^{ij}\widehat Q_i\widehat D_j\widehat H_d  
          +  h_E^{ij}\widehat L_i\widehat E_j\widehat H_d \nonumber
\\ 
        & + & h_{\nu}^{i}\widehat L_i\widehat \nu^c\widehat H_u 
          - h_0 \widehat H_d \widehat H_u \widehat\Phi 
          + h \widehat\Phi \widehat\nu^c\widehat S +
          \frac{\lambda}{3!} \widehat\Phi^3 .
\label{eq:Wsuppot} 
\end{eqnarray}
The first three terms are the usual MSSM Yukawa terms. The terms 
coupling the lepton doublets to $\widehat\nu^c$ fix lepton number. 
The coupling of the field $\widehat\Phi$ with the Higgs doublets 
generates an effective $\mu$-term a l\'a Next to Minimal Supersymmetric 
Standard Model \cite{nmssm}. The last two terms, involving 
only singlet fields, give mass to $\widehat\nu^c$, $\widehat S$ and 
$\widehat\Phi$, once $\Phi$ develops a vev. 

For simplicity we consider only one generation of $\widehat\nu^c$ and 
$\widehat S$. Adding more generations of $\widehat\nu^c$ or $\widehat S$ 
does not add any qualitatively new features to the model. Note also, 
that the superpotential, Eq.~(\ref{eq:Wsuppot}), does not contain any 
terms with dimension of mass, thus potentially offering a solution 
to the $\mu$-problem of supersymmetry. The soft supersymmetry breaking 
terms of this model can be found in \cite{Hirsch:2004rw}.

The basic guiding principle in the construction of Eq.~(\ref{eq:Wsuppot}) 
is that lepton number is conserved at the level of the superpotential. 
However, at low energy, i.e.~after electro-weak symmetry breaking, various 
fields acquire vevs. Besides the usual MSSM Higgs boson vevs $v_D$ and 
$v_U$, these are $\langle \Phi \rangle = v_P/\sqrt{2}$, 
$\langle {\tilde \nu}^c \rangle = v_R/\sqrt{2}$, 
$\langle {\tilde S} \rangle = v_S/\sqrt{2}$ and 
$\langle {\tilde \nu}_i \rangle = v_{L_i}/\sqrt{2}$. 
Note, that $v_R \ne 0$ generates effective bilinear terms 
$\epsilon_i = h_{\nu}^i v_R/\sqrt{2}$ and that $v_R$, $v_S$ and $v_{L_i}$ 
violate lepton number and R-parity spontaneously. 

Once R-parity is broken, Majorana neutrino masses are generated. 
After some algebraic manipulations one can express the effective 
(3,3) neutrino mass matrix in seesaw approximation as \cite{Hirsch:2004rw}
\begin{equation}
(\boldsymbol{m_{\nu\nu}^{\rm eff}})_{ij} = a \Lambda_i \Lambda_j + 
     b (\epsilon_i \Lambda_j + \epsilon_j \Lambda_i) +
     c \epsilon_i \epsilon_j\,,
\label{eq:eff}
\end{equation}
where $\Lambda_i = \epsilon_i v_d + \mu v_{L_i}$ and
$\mu = h_0 \frac{v_P}{\sqrt{2}}$. The coefficients $a$, $b$ and 
$c$ can be found in \cite{Hirsch:2004rw}. Equation (\ref{eq:eff}) resembles 
closely the structure found for the neutrino mass matrix in the explicit 
bilinear model at the one-loop level \cite{hirsch:2000ef}. However, 
with the field content of Eq.~(\ref{eq:Wsuppot}) the spontaneous model 
produces two non-zero neutrino masses already at tree-level. Note, that 
the coefficients $b$ and $c$ go to zero (and thus one of the two neutrino 
masses vanishes at tree-level) in the limit when either $h$  {\em and} 
$h_0$ vanish or, equivalently, $v_P \rightarrow \infty$ {\em or} 
$v_R\rightarrow  \infty$. 

Neutrino physics puts a number of constraints on the parameters $\Lambda_i$ 
and $\epsilon_i$. For the current paper the exact details are unimportant, 
however, the most essential constraint for the following discussion is that 
both, $\Lambda_i/m_{\rm SUSY}^2$ and $|\epsilon_i/\mu|$, are small. 
If some of the singlet fields are light, i.e.~have masses in the range 
of ${\cal O}(0.1-{\rm few})$ TeV, $|\epsilon_i/\mu|$ can be as small as 
$|\epsilon_i/\mu| \sim 10^{-6}$--$10^{-5}$ (implying $v_{L_i}/v_D \sim 
10^{-6}$--$10^{-4}$). On the other extreme, independent of the singlet 
spectrum, $|\epsilon_i/\mu|$ can not be larger than, say, $|\epsilon_i/\mu| 
\sim 10^{-3}$, due to contributions from sbottom 
and stau loops to the neutrino mass matrix \cite{hirsch:2000ef}. 

The requirement that $v_{L_i} \ll v_D$, together with the assumption 
of $v_{L_i}/V < 1$,  can be used to find a rather simple expression 
for the majoron, in the basis
$A'_0= Im(H_d^{0},H_u^{0},\tilde{\nu}_{1},\tilde{\nu}_{2},
\tilde{\nu}_{3},\Phi,\tilde{S},\tilde{\nu}^{c})$:
\begin{equation}
\label{eq:majntl}
J \simeq  (\frac{-v_d v_L^2}{Vv^2},\frac{v_u v_L^2}{Vv^2},
           \frac{v_{L1}}{V},\frac{v_{L2}}{V},\frac{v_{L3}}{V},
           0,\frac{v_S}{V},-\frac{v_R}{V})\,,
\end{equation}
where $ V^2=v_S^2+v_R^2$. We stress that neutrino data ensures 
$v_{L_i}/V < 1$ even for $v_R$ as small as ${\cal O}(0.1)$ GeV. 
Thus, the majoron in our model is mainly a gauge singlet. Constraints 
from the measurements of the invisible width of the $Z^0$ boson \cite{lep} 
as well as all astrophysical bounds \cite{Raffelt:1996wa} are therefore 
easily satisfied. Note, that the majoron has no component in the $\Phi$ 
direction. 

The coupling between the lightest neutralino, the neutrinos and the 
majoron in general is a complicated function of singlet and doublet 
parameters. However, in the limit $v_R, v_S \rightarrow \infty$ one 
can derive a simple approximation formula, given by
\begin{equation}
\label{eq:majcl}
C_{\tilde\chi^0_1\nu_kJ} \simeq  \frac{{\tilde \epsilon}_k}{V}N_{14} +
\frac{{\tilde v}_{L_k}}{V}(g' N_{11} - g N_{12})+ \cdots
\end{equation}
Here, ${\tilde \epsilon}= U_{\nu}^T \cdot {\vec \epsilon}$ and 
 ${\tilde v_L}= U_{\nu}^T \cdot {\vec v_L}$, where $U_{\nu}^T$ 
is the matrix which diagonalizes the neutrino mass matrix 
at tree-level. The dots in 
Eq.~(\ref{eq:majcl}) stand for higher order terms. Equation(\ref{eq:majcl}) 
serves to show that for constant ${\tilde \epsilon}$ and 
${\tilde v}_{L}$, $C_{\tilde\chi^0_1\nu_kJ} \rightarrow 0$ as $v_R$ 
goes to infinity. This is as expected, since for $v_R\rightarrow\infty$ 
the spontaneous model approaches the explicit bilinear model. 
Note, that only the presence of the field $\widehat \nu^c$ is essential 
for the coupling Eq.~(\ref{eq:majcl}). If $\widehat S$ is absent, 
replace $V \rightarrow v_R$. 

We now turn to the numerical results. All calculations presented below 
are done with the program package SPheno \cite{Porod:2003um}, extended 
to include R-parity violation. We assume that the singlet states of the 
model are too heavy to be produced at future collider experiments such 
that only MSSM particles are observed. Note, that the smallness 
of the \rpv parameters guarantees that R-parity violating decay modes of 
all supersymmetric particles can be neglected except for the LSP and that 
also SUSY particle production cross sections are completely MSSM-like. 
We will study only points where the lightest neutralino $\tilde \chi^0_1$
is the LSP.

\begin{figure}[t]
\begin{center}
\vspace{5mm}
\includegraphics[width=85mm,height=60mm]{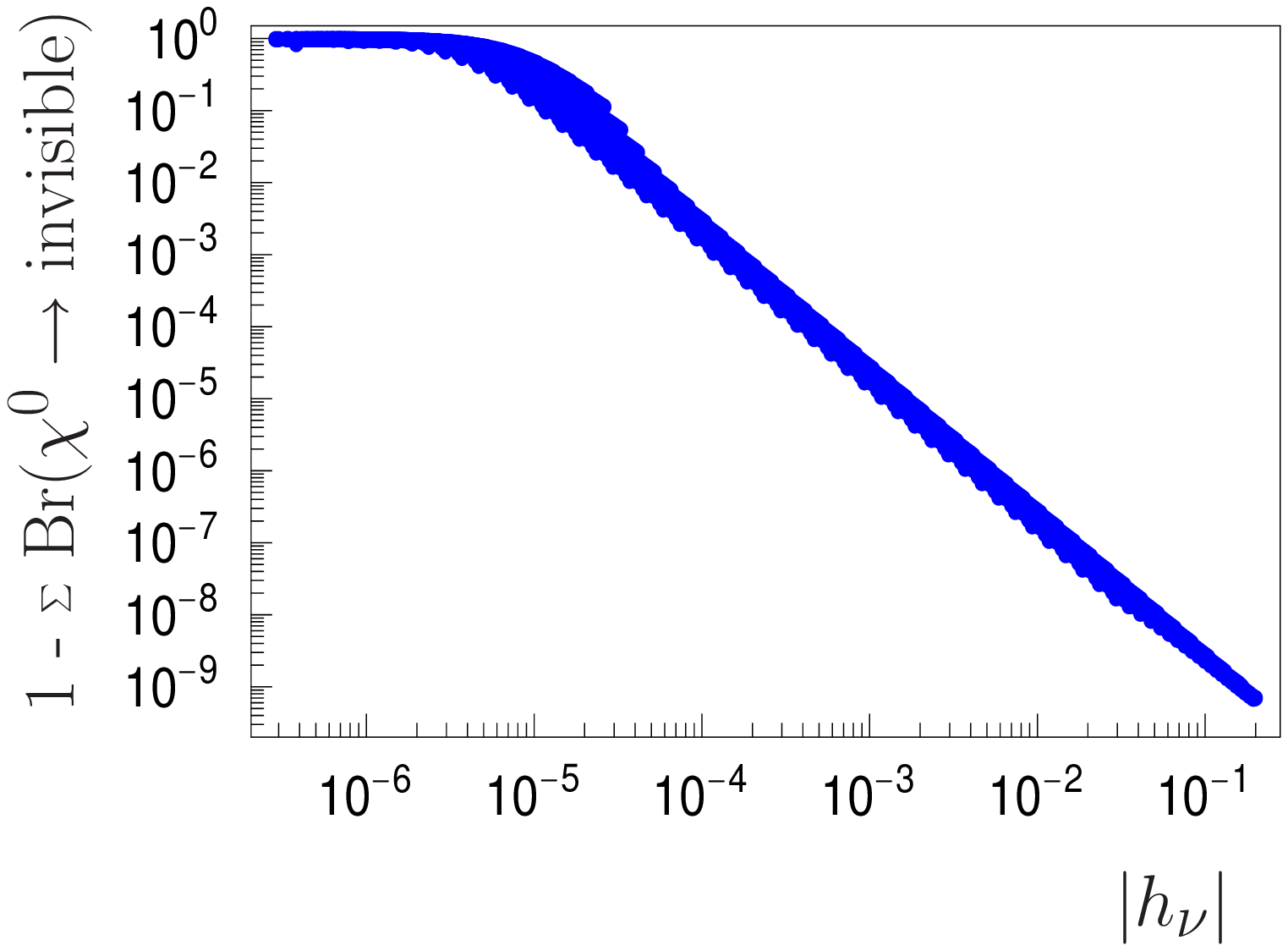}
\includegraphics[width=85mm,height=60mm]{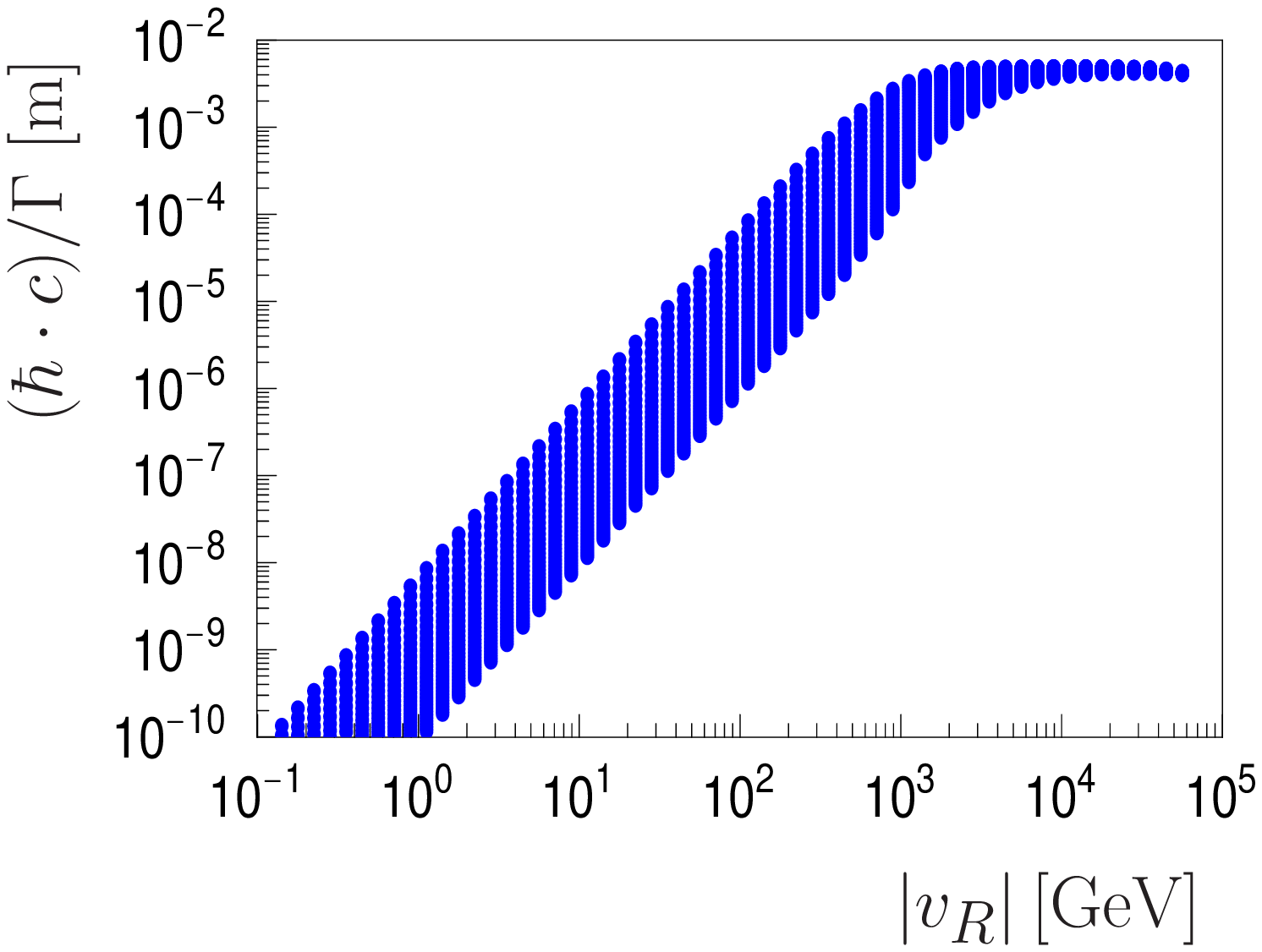}
\end{center}
\vskip-120mm\hskip115mm (a)

\vskip55mm\hskip65mm (b)

\vspace{48mm}
\caption{(a) Sum over all at least partially visible decay modes 
of the lightest neutralino versus $|h^{\nu}| = \sqrt{\sum_i(h_{\nu}^i)^2}$ 
for mSUGRA parameter point SPS1a' \cite{Aguilar-Saavedra:2005pw}. 
(b) Same points, decay length versus $v_R$. This is the main result of 
the current paper.}
\label{fig:SumVis}
\end{figure}

As can be seen from Eq.~(\ref{eq:majcl}), adjusting $\epsilon_i$ and 
$v_{L_i}$ from neutrino data, see Eq.~(\ref{eq:eff}), one expects that 
the most important parameters for the decay $\tilde\chi^0 \rightarrow J\nu$ 
are $|h^{\nu}| = \sqrt{\sum_i(h_{\nu}^i)^2}$ and, equivalently, $v_R$. 
Therefore, in Fig.~\ref{fig:SumVis}(a) we plot the sum over the branching 
ratios of neutralino decays containing at least one visible particle in 
the final state as a function of $|h^{\nu}|$, for fixed values of all 
other parameters except $v_P = 10$--40 TeV. For this plot all MSSM 
parameters were chosen in such a way that the sparticle spectrum as well 
as all production cross sections very closely resemble the values of the 
SPA point SPSA1a' \cite{Aguilar-Saavedra:2005pw}. For this point one expects 
about 2.5 $10^7$ neutralinos at the LHC for a total luminosity of 300 
fb$^{-1}$. As Fig.~\ref{fig:SumVis}(a) demonstrates, the total number of 
events with at least some visible energy in the final state drops 
in this example below 100 (10) for $|h^{\nu}| \le 2.5 \cdot 10^{-3}$ 
($7.5 \cdot 10^{-3}$). Note, that in any SUSY event there are two 
$\tilde \chi^0_1$ and that with such small branching ratios at most one 
of them decays visibly. Therefore, establishing that the $\tilde \chi^0_1$ 
decays at the LHC becomes challenging.

For completeness, in Fig.~\ref{fig:SumVis}(b) we plot $c\tau$ as a 
function of $v_R$ for the same points as in Fig.~\ref{fig:SumVis}(a). 
For SPS1a', the neutralino decays always with $L \le 1$ cm. Other 
points have different upper limits, but $L \le 1$ m always. 

We have explicitly checked that choosing different MSSM parameters reproduces 
Fig.~\ref{fig:SumVis}(a) for a number of standard points from refs. 
\cite{Battaglia:2003ab,DeRoeck:2005bw}, as well as for points with 
a calculated $\Omega_{\tilde\chi^0} \simeq 1$ (in case $R_p$ were conserved). 
The invisible decay width of the neutralino does essentially not depend 
on MSSM parameters and thus is independent of $\Omega_{DM}$ obtained in 
conserved R-parity. 

\begin{figure}[t]
\begin{center}
\vspace{5mm}
\includegraphics[width=80mm,height=55mm]{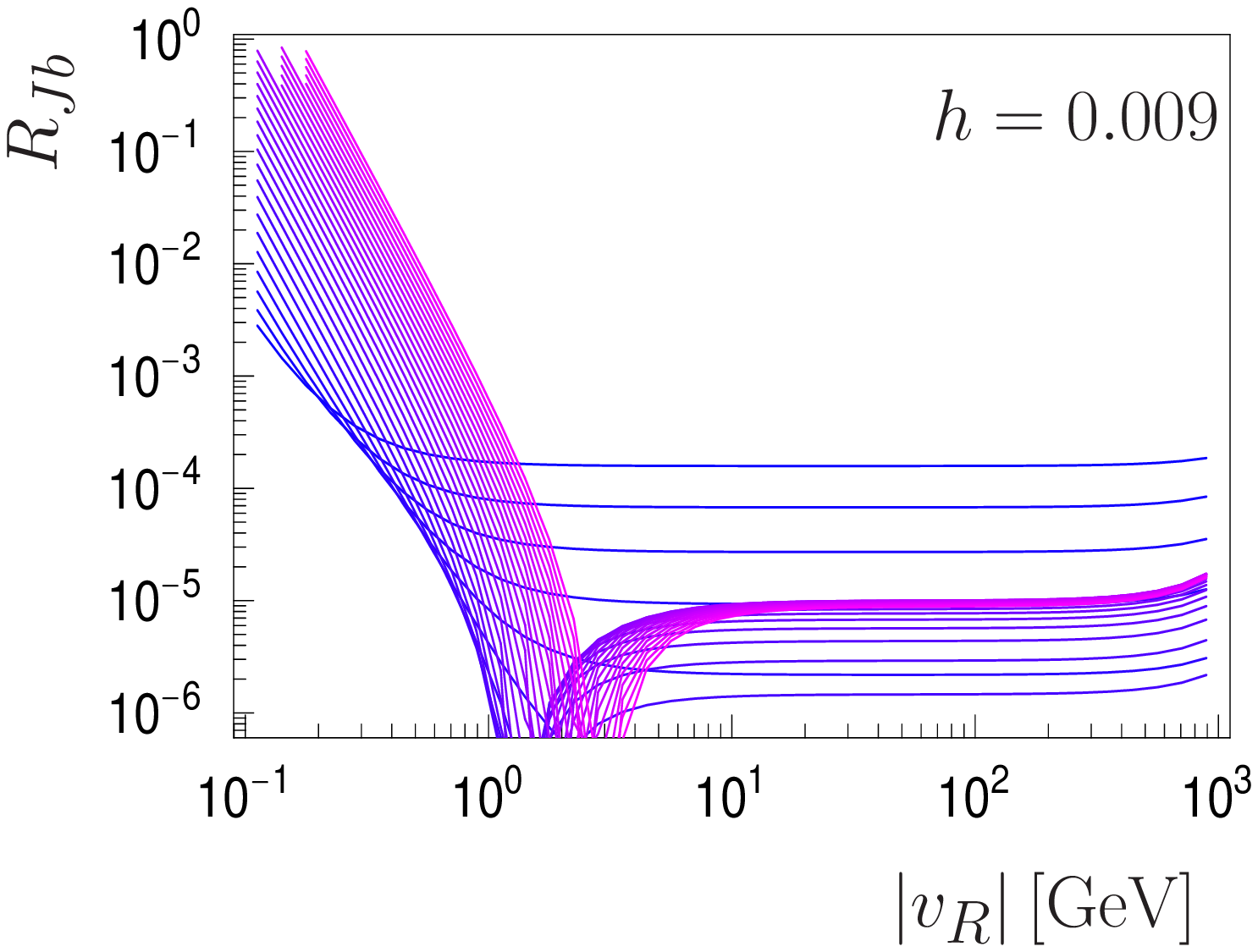}
\includegraphics[width=80mm,height=55mm]{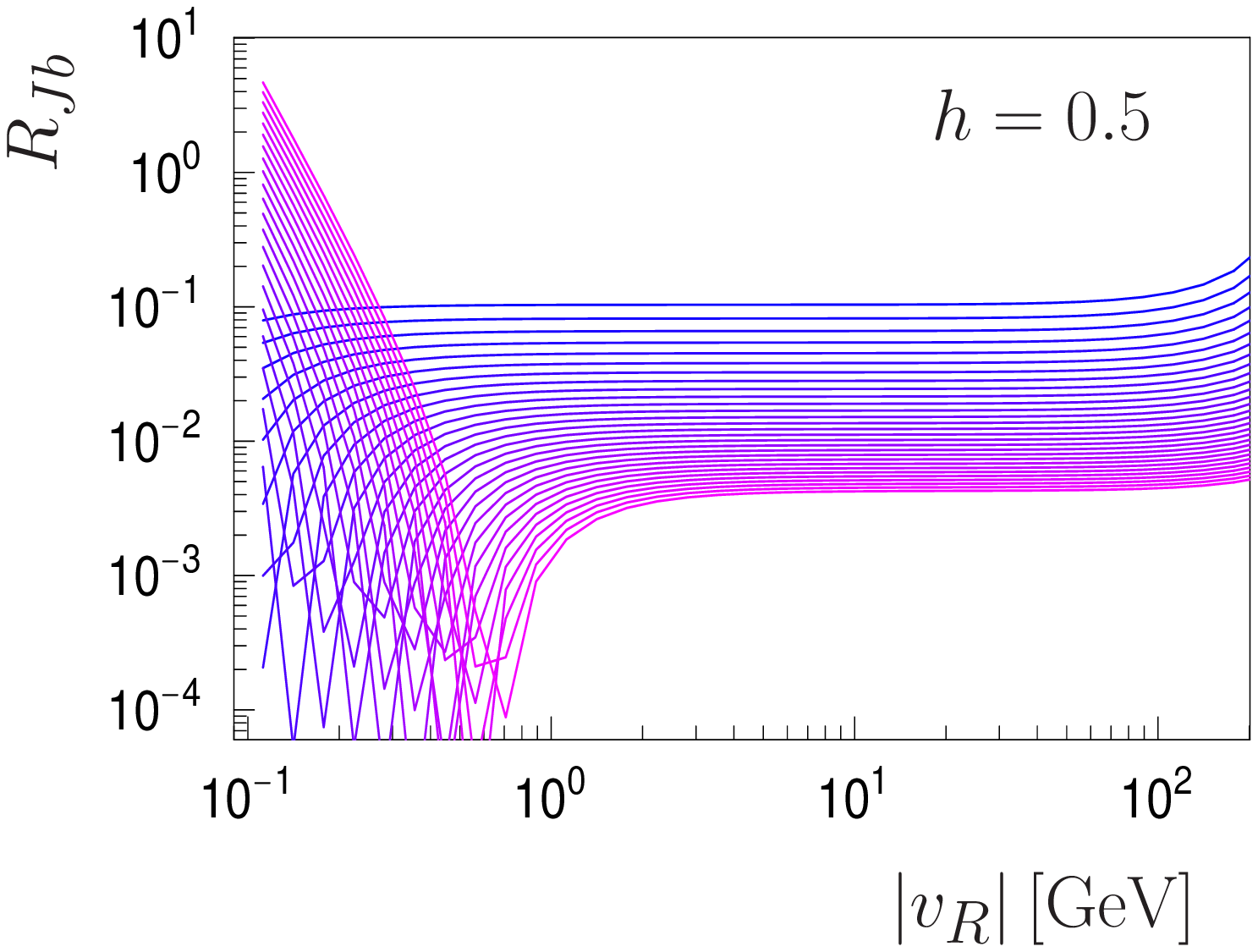}
\end{center}
\vspace{-5mm}
\caption{$R_{Jb}$ defined as $\Gamma(S_{h^0} \rightarrow JJ)/ 
\Gamma(S_{h^0} \rightarrow b{\bar b})$ as a function of $v_R$ 
for the mSUGRA point SPS1a' \cite{Aguilar-Saavedra:2005pw} 
for two values of the coupling $h$, for a number of 
different values of $v_P$, $v_P = 10$ TeV (blue) to $v_P = 40$ TeV 
(red) in steps of $1$ TeV.}
\label{fig:HiggsInvis}
\end{figure}

For this model to resemble the MSSM it is not only important that the
SUSY decays are MSSM-like and that the LSP decays invisibly but also
that the Higgs sector behaves as in the MSSM. This is potentially in
conflict with an interesting feature of this model, namely the invisible
decay of the lightest doublet Higgs boson into two Majorons
\cite{Hirsch:2004rw}. In Fig.~\ref{fig:HiggsInvis} we show the ratio 
$R_{Jb}=\Gamma(S_{h^0} \rightarrow JJ)/ \Gamma(S_{h^0}\rightarrow b{\bar b})$ 
for the same points as shown in Fig.~\ref{fig:SumVis}, i.e.~$h=0.009$, 
as well as for a different choice of this parameter, $h=0.5$. Note that 
the latter choice leads to invisible $\tilde \chi^0$ decay branching ratios 
indistinguishable from Fig.~\ref{fig:SumVis}. Obviously there is enough 
freedom in choosing the parameters such that the invisible decay of the 
Higgs varies over several orders of magnitude, from a completely negligible 
level up to branching ratios similar to the dominant $b \bar{b}$ mode. 
We mention that the observed 'dips' in $R_{Jb}$  are due to a cancellation 
between various contributions in the $C_{h^0JJ}$ coupling. 
Figure~\ref{fig:HiggsInvis} demonstrates that not only the properties of 
the SUSY particles can be adjusted to be MSSM-like (with conserved R-parity) 
but also the Higgs sector. 

A final comment: We have chosen s-\rpv parameters fitted to neutrino masses. 
Smaller parameters $\vec\epsilon$ and $\vec v_L$ lead to too small neutrino 
masses, but are allowed phenomenologically. These would lead to larger 
decay lengths, but not to larger branching ratios into visible final 
states \footnote{Adding some trilinear coupling, essentially 
unconstrained by neutrino masses \cite{Rakshit:1998kd}, like $\lambda_{121}$ 
would of course lead to a larger number of visible events.}.

In summary, we have studied the decay properties of a neutralino LSP in 
a model with spontaneous violation of R-parity. The neutralino decays 
invisibly in this model in large parts of the parameter space, such that 
even after several years of data taking at LHC it might be experimentally 
challenging to establish that R-parity is broken. A first hint of a 
non-standard scenario might come from cosmology, as in most of MSSM 
parameter space the relic density of the neutralinos is too large to 
be compatible with astronomical observations. However, with sufficiently 
large statistics it should eventually be possible to discover the rare 
visible decays of the lightest neutralino, e.g. into $l^\pm q \bar{q}'$ 
final states, and thus prove that R-parity is violated. Needless to say, 
that such a search requires a dedicated effort by the experimentalists.

\section*{Acknowledgments}
We thank R.~Tomas for discussion on supernova and the derivation of bounds
on model parameters. This work was supported by Spanish grant FPA2005-01269, 
by the European Commission Human Potential Program RTN network
MRTN-CT-2004-503369.  M.H. and W.P. are supported by MCyT Ramon y Cajal 
contracts.

\end{document}